\documentclass[aps,prd,twocolumn,showpacs,nofootinbib]{revtex4-1}

\usepackage{graphicx}
\usepackage{amssymb}
\usepackage{color}

\usepackage[colorlinks=true,linkcolor=blue,citecolor=blue, urlcolor=blue]{hyperref}

\begin{document}

\title{QCD corrections to the $B_c$ to charmonia semi-leptonic decays}

\author{Jian-Ming Shen}
\author{Xing-Gang Wu} \email{wuxg@cqu.edu.cn}
\author{Hong-Hao Ma}
\author{Sheng-Quan Wang}

\affiliation{Department of Physics, Chongqing University, Chongqing 401331, P.R. China}

\date{\today}

\begin{abstract}
We present a detailed analysis on the $B_c$ meson semi-leptonic decays, $B_c \to \eta_c (J/\psi) \ell \nu$, up to next-to-leading order (NLO) QCD correction. We adopt the principle of maximum conformality (PMC) to set the renormalization scales for those decays. After applying the PMC scale setting, we determine the optimal renormalization scales for the $B_c\to\eta_c(J/\psi)$ transition form factors (TFFs). Because of the same $\beta_0$-terms, the optimal PMC scales at the NLO level are the same for all those TFFs, i.e. $\mu_r^{\rm PMC} \approx 0.8{\rm GeV}$. We adopt a strong coupling model from the massive perturbation theory (MPT) to achieve a reliable pQCD estimation in this low energy region. Furthermore, we adopt a monopole form as an extrapolation for the $B_c\to\eta_c(J/\psi)$ TFFs to all their allowable $q^2$ region. Then, we predict $\Gamma_{B_c \to \eta_c \ell \nu}(\ell=e,\mu) =(71.53^{+11.27}_{-8.90})\times 10^{-15} {\rm GeV}$, $\Gamma_{B_c \to \eta_c \tau \nu}=(27.14^{+5.93}_{-4.33})\times 10^{-15} {\rm GeV}$, $\Gamma_{B_c \to J/\psi \ell \nu}(\ell=e,\mu) =(106.31^{+18.59}_{-14.01}) \times 10^{-15} {\rm GeV}$, $\Gamma_{B_c \to J/\psi \tau \nu} =(28.25^{+6.02}_{-4.35})\times 10^{-15} {\rm GeV}$, where the uncertainties are squared averages of all the mentioned error sources. We show that the present prediction of the production cross section times branching ratio for $B^+_c\to J/\psi \ell^+ v$ relative to that for $B^+ \to J/\psi K^+$, i.e. $\Re(J/\psi \ell^+ \nu)$, is in a better agreement with CDF measurements than the previous predictions.
\end{abstract}


\maketitle

\section{Introduction}
\label{sec:1}

The $B_c$ meson is a doubly heavy quark-antiquark system and carries flavors explicitly. It has been pointed out that sizable $B_c$ meson events can be produced at the hadronic colliders~\cite{bc1,bc2,bc3}. Thus, it provides a useful laboratory for studying both the Quantum Chromodynamics (QCD) and the weak interaction theories.

\begin{table}[htb]
\begin{tabular}{c|c}
\hline\hline
    & ~~$\Re (J/\psi \ell^+ \nu)$~~\\
\hline
$p_{\rm T}>6{\rm GeV}$~\cite{Bc98} & $0.132_{-0.037}^{+0.041} ({\rm st}) \pm 0.031({\rm sy}) _{-0.020}^{+0.032}({\rm lf})$ \\
$p_{\rm T}>4{\rm GeV}$~\cite{Ru05} & $0.249 \pm 0.045 ({\rm st}) \pm 0.069 ({\rm sy}) _{ - 0.033}^{ + 0.082}({\rm lf})$ \\
$p_{\rm T}>6{\rm GeV}$~\cite{Ru05} & $0.245 \pm 0.045 ({\rm st}) \pm 0.066 ({\rm sy}) _{ - 0.032}^{ + 0.080}({\rm lf})$ \\
$p_{\rm T}>4{\rm GeV}$~\cite{Re05} & $0.282 \pm 0.038 ({\rm st}) \pm 0.035 ({\rm y}) \pm 0.065 ({\rm a})$ \\
$p_{\rm T}>6{\rm GeV}$~\cite{Re05} & $0.242 \pm 0.036 ({\rm st}) \pm 0.031 ({\rm y}) \pm 0.051 ({\rm a})$ \\
$p_{\rm T}>4{\rm GeV}$~\cite{Ru09} & $0.295 \pm 0.040 ({\rm st}) _{ - 0.026}^{ + 0.033} ({\rm sy}) \pm 0.036 ({\rm sp})$ \\
$p_{\rm T}>6{\rm GeV}$~\cite{Ru09} & $0.227 \pm 0.033 ({\rm st}) _{ - 0.017}^{ + 0.024} ({\rm sy}) \pm 0.014 ({\rm sp})$ \\
$p_{\rm T}>6{\rm GeV}$~\cite{Ru14} & $0.211 \pm 0.012 ({\rm st}) _{ - 0.020}^{ + 0.021} ({\rm sy})$ \\
\hline\hline
\end{tabular}
\caption{The $\Re (J/\psi \ell^+ \nu)$ measured by CDF collaboration under two $p_T$ cuts. The symbols ``st'', ``sy'', ``lf'', ``y'', ``a'' and ``sp'' stand for the statistical error, the systematic error, the $B_c$ lifetime error, the systematic error on the yield, the systematic error on acceptance ratio and the $B_c$ spectrum error, respectively. The second line is for  $\Re (J/\psi e^+ \nu)+\Re (J/\psi \mu^+ \nu)$, the fifth and sixth lines are for $\Re (J/\psi e^+ \nu)$, and the remaining lines are for $\Re (J/\psi \mu^+ \nu)$.}
\label{tab:Rexpe}
\end{table}

Experimentally, the Collider Detector at Fermilab (CDF) collaboration discovered the $B_c$ meson in the year 1998 via the semi-leptonic decay channel $B_c^+ \to J/\psi \ell^+ \nu$~\cite{Bc98}. It also predicted the value of the production cross section times branching ratio fraction between the $B_c^+ \to J/\psi \ell^+ \nu$ and $B^+ \to J/\psi K^+$, i.e.
\begin{equation}
\Re(J/\psi \ell^+ \nu) = \frac{\sigma(B_c^+) BR(B_c^+ \to J/\psi \ell^+ \nu)}{\sigma(B^+) BR(B^+ \to J/\psi K^+)}. \label{ration}
\end{equation}
Later on, more measurements for the $B_c$ meson properties have been done at both the Tevatron and the LHC colliders, cf.Refs.\cite{CDFexp08, D0exp08, D0exp09, LHCbexp1,LHCbexp2,LHCbexp3,LHCbexp4,LHCbexp5,LHCbexp6, Ru05,Re05, Re06, RuRe05, Ru09, Ru14}. For our present purpose, we put the values of $\Re(J/\psi \ell^+ \nu)$ predicted by the CDF collaboration in Table \ref{tab:Rexpe}, in which two $B_c$ meson $p_T$ cuts have been adopted.

Theoretically, the predicted values for $\Re(J/\psi{\ell^+}\nu)$ are always smaller than the experimental measurements. Such a comparison has firstly been done by Ref.\cite{Bc98}, whose Fig.(3) shows that the theoretical predictions are well below the CDF prediction. As shown by Table~\ref{tab:Rexpe}, the updated Tevatron Run II measurements for $\Re(J/\psi{\ell^+}\nu)$~\cite{Ru05,Re05,Re06,RuRe05,Ru09,Ru14} are almost doubled in comparison to the previous one~\cite{Bc98}, then the discrepancy becomes even worse. This discrepancy arouses people's great interests, many attempts have been tried to solve the puzzle. It has been argued that, by including the next-to-leading order (NLO) QCD correction to the $B_c\to\eta_c(J/\psi)$ semi-leptonic decays, the prediction on $\Re(J/\psi{\ell^+}\nu)$ can be consistent with the experimental results~\cite{qiao12,qiao13}. However, in Ref.\cite{qiao12} the NLO estimation is done for a fixed $\alpha_s\simeq0.2$ and in Ref.\cite{qiao13}, a quite small $\Lambda_{\rm QCD}\simeq 0.1$ GeV has been adopted. Even worse, a large renormalization scale uncertainty and a large NLO contributions (or a large $K$ factor) make the pQCD prediction questionable.

It is noted that the pQCD series in Refs.~\cite{qiao12,qiao13} is derived under the conventional scale setting, in which the renormalization scale is set to be the typical momentum flow ($Q\sim m_b$) of the process and an arbitrary region as $[Q/2,2Q]$ is adopted for estimating the scale error. The conventional scale setting assigns an arbitrary range and an arbitrary systematic error to the fixed-order pQCD prediction~\cite{PMC3}, thus, it is natural to assume that the present puzzling situation may not be the question of the pQCD theory but the question of the conventional scale setting. As an attempt to improve the pQCD predictions, we shall apply the principle of maximum conformality (PMC)~\cite{PMC1,PMC2,PMC3,PMC4} to deal with $B_c\to\eta_c(J/\psi)$ semi-leptonic decays up to NLO level.

The PMC provides a systematic procedure to set the renormalization scale for any QCD processes. It is well-known that the running behavior of the coupling constant is governed by the renormalization group equation (RGE). Following such principal, the PMC is to use the $\beta$-terms in the perturbative series to determine the optimal behavior of coupling constant, or equivalently, to determine the optimal scale for the coupling constant~\cite{PMC4}. One can start the pQCD calculation with an arbitrary but hard enough initial scale. Then, by finding out all the related $\beta$-terms that rightly determine the running behavior of coupling constant at each perturbative order and resuming them into the coupling constant, the argument of the coupling constant at each perturbative order shall be shifted from its initial value to its optimal one. The resultant PMC expressions being free of $\beta$-terms are thus independent of the renormalization scheme, as required by the renormalization group invariance~\cite{PMC5}. In the present paper, we shall show that after applying the PMC scale setting, an improved QCD estimation for $B_c \to \eta_c(J/\psi) \ell \nu$ can be achieved, i.e. the scale uncertainty can be greatly suppressed and a more reasonable central value for the decay width of $B_c \to \eta_c(J/\psi) \ell \nu$ can be achieved.

As the key components for the $B_c \to \eta_c(J/\psi) \ell \nu$ semi-leptonic decays, the $B_c \to \eta_c(J/\psi)$ transition form factors (TFFs) have been calculated up to NLO level. It is noted that those TFFs are pQCD calculable only in large recoil region with $q^2\sim0$. Thus, one needs to extrapolate them to all their allowable physical region so as to estimate the total decay widths (or the branching ratios) of $B_c \to \eta_c(J/\psi) \ell \nu$. Several extrapolation approaches have been suggested in the literature, cf. Refs.\cite{Kiselev00,Kiselev02, Huang07,Lu09,BSWFF85,monopole}. In the present work, we shall adopt the monopole form to do the extrapolation, which has been firstly suggested in Ref.\cite{monopole}. Then, we shall reestimate the value of $\Re(J/\psi \ell^+ \nu)$ and make a comparison with the CDF predictions.

The remaining parts of the paper are organized as follows. In Sec.\ref{sec:2}, we present the calculation technology for dealing with the $B_c \to \eta_c(J/\psi)$ TFFs up to NLO level. The $B_c \to \eta_c(J/\psi)$ TFFs and their relations to the $B_c$ meson semi-leptonic decay widths are presented. The PMC treatment and the treatment of the low-energy running coupling are also presented here. In Sec.\ref{sec:3}, we present the numerical results for the TFFs at the large recoil region, and the $\Re(J/\psi{\ell^+}\nu)$ is recalculated and compared with the experimental predictions. The last section is reserved for a summary.

\section{Calculation technology}
\label{sec:2}

After integrating the phase-space, the differential decay width over $q^2$ for the semi-leptonic decay $B_c \to \eta_c \ell \nu$ or $B_c \to J/\psi \ell \nu$ can be formulated as
\begin{widetext}
\begin{eqnarray}
\frac{{\rm d}\Gamma(B_c\to\eta_c\ell\nu)}{{\rm d}q^2} &=& \bigg(\frac{q^2-m_{\ell}^2}{q^2}\bigg)^2 \frac{\sqrt{\lambda_1(q^2)} G_F^2 |V_{\rm cb}|^2}{384m_{B_c}^3\pi^3 q^2} \left[ (m_{\ell}^2+2q^2) \lambda_1(q^2) F_1^2(q^2)+3m_{\ell}^2(m_{B_c}^2-m_{\eta_c}^2)^2 F_0^2(q^2) \right], \\
\frac{{\rm d}\Gamma_{\rm L}(B_c\to J/\psi\ell\nu)}{{\rm d}q^2} &=& \bigg(\frac{q^2-m_{\ell}^2}{q^2}\bigg)^2 \frac{\sqrt{\lambda_2(q^2)} G_F^2 |V_{\rm cb}|^2} {384m_{B_c}^3\pi^3} \left[\frac{3m_{\ell}^2}{q^2} \lambda_2(q^2) A_0^2(q^2)+(m_{\ell}^2+2q^2)|h_0(q^2)|^2 \right], \\
\frac{{\rm d}\Gamma_{\rm T}(B_c\to J/\psi\ell\nu)}{{\rm d}q^2} &=& \bigg(\frac{q^2-m_{\ell}^2}{q^2}\bigg)^2 \frac{\sqrt{\lambda_2(q^2)} G_F^2 |V_{\rm cb}|^2} {384m_{B_c}^3\pi^3} \times (m_{\ell}^2+2q^2) \left[|h_+(q^2)|^2 + |h_-(q^2)|^2 \right],
\end{eqnarray}
\end{widetext}
where we have separated the total decay for the $J/\psi$ case as $\Gamma=\Gamma_{\rm L}+ \Gamma_{\rm T}$, the lepton $\ell=e,\mu,\tau$ and the Fermi constant $G_F=1.16638\times10^{-5}$~\cite{RPP}. $q=P-p$ is the momentum transfer, $P$ is momentum of $B_c$ meson and $p$ is momentum of $\eta_c$ or $J/\psi$. The phase-space factors $\lambda_1(q^2)=(m_{B_c}^2+m_{\eta_c}^2-q^2)^2- 4 m_{B_c}^2 m_{\eta_c}^2$ and $\lambda_2(q^2)=(m_{B_c}^2+m_{J/\psi}^2-q^2)^2-4m_{B_c}^2m_{J/\psi}^2$. The longitudinal and transverse helicity amplitudes for $\Gamma_{\rm L}$ and $\Gamma_{\rm T}$ are expressed as:
\begin{widetext}
\begin{eqnarray}
h_{\pm}(q^2) &=& \frac{\sqrt{\lambda_2(q^2)}}{m_{B_c}+m_{J/\psi}}\Big[V(q^2) \mp \frac{(m_{B_c}+m_{J/\psi})^2}{\sqrt{\lambda_2(q^2)}}A_1(q^2)\Big],  \\
h_0(q^2) &=& \frac{1}{2m_{J/\psi}\sqrt{q^2}} \Big[-\frac{\lambda_2(q^2)}{m_{B_c}+m_{J/\psi}}A_2(q^2) +(m_{B_c}+m_{J/\psi})(m_{B_c}^2-m_{J/\psi}^2-q^2)A_1(q^2)\Big].
\end{eqnarray}
\end{widetext}
The two $B_c\to\eta_c$ TFFs $F_0(q^2)$ and $F_1(q^2)$, and the four $B_c\to J/\psi$ TFFs $V(q^2)$, $A_0(q^2)$, $A_1(q^2)$ and $A_2(q^2)$, are defined as follows~\cite{BSWFF85}:
\begin{widetext}
\begin{eqnarray}
\langle \eta_c(p) |\bar c \gamma^{\mu} b| B_c(P)\rangle &=& F_0(q^2)\frac{m_{B_c}^2-m_{\eta_c}^2}{q^2}q^{\mu} + F_1(q^2) \bigg(P^{\mu}+p^{\mu}-\frac{m_{B_c}^2-m_{\eta_c}^2}{q^2}q^{\mu} \bigg), \\
\langle J/\psi(p,\varepsilon^*) |\bar c \gamma^{\mu}b| B_c(P)\rangle
&=& \frac{2 i V(q^2)}{m_{B_c}+m_{J/\psi}} \epsilon^{\mu\nu\rho\sigma} \varepsilon_{\nu}^* p_{\rho}P_{\sigma}, \\
\langle J/\psi(p,\varepsilon^*) |\bar c \gamma^{\mu}{\gamma_5}b| B_c(P)\rangle &=& 2 m_{J/\psi}\frac{{\varepsilon^*} \cdot q}{q^2} q^{\mu} A_0(q^2) + (m_{B_c}+m_{J/\psi}) \bigg( \varepsilon^{*\mu}-\frac{\varepsilon^* \cdot q}{q^2} q^{\mu} \bigg) A_1(q^2) \nonumber\\
&& - \bigg({P^\mu} + {p^\mu} - \frac{m_{B_c}^2 - m_{J/\psi}^2}{q^2}q^{\mu}\bigg)\frac{{\varepsilon^*} \cdot q}{m_{B_c}+m_{J/\psi}} A_2(q^2).
\end{eqnarray}
\end{widetext}
For the case of $\ell=e$ or $\mu$, the lepton mass $m_\ell$ tends to zero, the contributions from $F_0$ and $A_0$ can be safely neglected due to the chiral suppression. All those TFFs are key components for the $B_c$ meson decays to charmonia. In the large recoil region, they are pQCD calculable and have been calculated up to NLO level~\cite{Bell06,Bell07,qiao12,qiao13}. Up to NLO level, we can schematically write the TFFs in the following form:
\begin{widetext}
\begin{equation}
f_i(q^2)=C^{f_i}(q^2)\alpha_s(\mu_r^{\rm init}) \left[1 + B^{f_i}(q^2,\mu_r^{\rm init}) \frac{\alpha_s(\mu_r^{\rm init}) }{\pi} \right], \label{starting}
\end{equation}
\end{widetext}
where $\mu_r^{\rm init}$ stands for some arbitrary initial renormalization scale, which should be large enough to ensure the pQCD calculation. For example, it can be chosen as the typical momentum flow of the process, i.e. $\mu_r^{\rm init}=m_b$. Under the conventional scale setting, the renormalization scale is fixed to be $\mu_r^{\rm init}$. On the other hand, for a certain scale setting, its resultant optimal scale depends on how we deal with the perturbative series and is usually different from $\mu_r^{\rm init}$. As for the PMC scale setting, its optimal scale is determined by the $\{\beta_i\}$-terms that rightly governs the running of the coupling constant via RGE. The function $f_i$ represents any one of the TFFs, $F_1(q^2)$, $F_0(q^2)$, $V(q^2)$, $A_0(q^2)$, $A_1(q^2)$ and $A_2(q^2)$, respectively. The tree-level coefficients $C^{f_i}(q^2)$ are put in the appendix and the NLO coefficients $B^{f_i}(q^2,\mu_r^{\rm init})$ can be read from Refs.\cite{Bell06,Bell07,qiao12,qiao13}. In those references, only the asymptotic expressions under the limit $m_b\to\infty$ have been given. Fortunately, however, as pointed out by Ref.\cite{qiao13}, those approximate expressions are of high precision in comparison to the full expressions~\footnote{We thank the authors of Ref.\cite{qiao13} for helpful discussions on this point and for kindly supplying us their Mathematical program for calculating the TFFs up to NLO level.}.

\subsection{The PMC scale setting for the TFFs}

\begin{figure}
\includegraphics[width=0.45\textwidth]{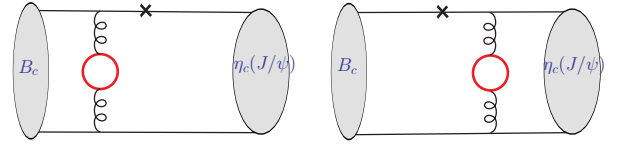}
\caption{Two NLO diagrams (together with their counter-terms) that contribute to the $\beta_0$-terms for the $B_c\to \eta_c (J/\psi)$ TFFs. The `cross' symbol means the weak interaction vertex and the circle stands for the light-quark loop. } \label{Feyn}
\end{figure}

To set the PMC scales for the TFFs, $F_1(q^2)$, $F_0(q^2)$, $V(q^2)$, $A_0(q^2)$, $A_1(q^2)$ and $A_2(q^2)$, we first decompose the NLO coefficients $B^{f_i}(q^2,\mu_r^{\rm init})$ into the non-conformal $\{\beta_i\}$-dependent part and the conformal $\{\beta_i\}$-independent part, i.e.
\begin{equation}
B^{f_i}(q^2,\mu_r^{\rm init}) = B^{f_i}_{(\beta)}(q^2,\mu_r^{\rm init}) \beta_0 + B^{f_i}_{\rm conf}(q^2,\mu_r^{\rm init}),
\end{equation}
where $\beta_0=11-\frac{2}{3}{n_f}$. The conformal part $B^{f_i}_{\rm conf}(q^2,\mu_r^{\rm init})$ are in general complex and their analytic expressions at $q^2=0$ are put in the Appendix. The coefficients of non-conformal part are the same for all TFFs, i.e.
\begin{eqnarray}
B^{f_i}_{(\beta)}(q^2,\mu_r^{\rm init}) = \frac{5}{12} + \frac{1}{4} \ln \left( \frac{(\mu_r^{\rm init})^2}{2\gamma m_c^2} \right), \label{betaterm}
\end{eqnarray}
where $\gamma=\frac{m_b^2-q^2}{4 m_b m_c}$ is the relativistic boost. One may observe that at the NLO level, all the TFFs have the same $n_f$-terms and hence the same $\beta_0$-terms. As shown by Fig.(\ref{Feyn}), this is due to the same one-loop gluon self-energy diagrams (together with their counter-term ones) for all the TFFs. Such $\beta_0$-terms rightly determine the running behavior of the LO coupling constant, thus, they should be absorbed into the coupling constant following the RGE~\cite{PMC1,PMC2,PMC3,PMC4}.

More specifically, after applying the PMC scale setting, the TFFs (\ref{starting}) shall be transformed as
\begin{widetext}
\begin{eqnarray}
f_i^{\rm PMC}(q^2) = C^{f_i}(q^2) \alpha_s(\mu_r^{\rm PMC})\left[ 1 + \frac{\alpha_s(\mu_r^{\rm PMC})}{\pi}B_{\rm conf}^{f_i}(q^2,\mu_r^{\rm init}) \right], \label{startingPMC}
\end{eqnarray}
\end{widetext}
where to eliminate the non-conformal $\beta_0$-terms, the renormalization scale has been transformed from its initial value $\mu^{\rm init}_r$ to the LO PMC scale $\mu_r^{\rm PMC}$, i.e.
\begin{eqnarray}
\mu_r^{\rm PMC} &=& \mu_r^{\rm init} \exp{\left(-2B_{(\beta)}^{f_i}(q^2,\mu_r^{\rm init})\right)} .
\end{eqnarray}
It is noted that the scale displacement contains a similar function as the simplest scale displacement $e^{-{5}/{6}}$ that ensures the scheme invariance between the $\overline{\rm MS}$ scheme and the Gell-Mann-Low scheme~\cite{gml}. More explicitly, with the help of Eq.(\ref{betaterm}), the LO PMC scale can be simplified as
\begin{equation}
\mu_r^{\rm PMC} = e^{-\frac{5}{6}}\sqrt{\frac{m_c}{2m_b}(m_b^2-q^2)}, \label{pmcscale}
\end{equation}
The interesting point is that the LO PMC scale is independent of $\mu^{\rm init}_r$. As discussed above, since all the TFFs has the same $\beta_0$-term $\left[B^{f_i}_{(\beta)}(q^2,\mu_r^{\rm init}) \beta_0\right]$, all of them shall have the same PMC scale. We have no NNLO $\{\beta_i\}$-terms to set the NLO PMC scale for $\alpha^2_s$-terms, and we have implicitly set $\mu_r^{\rm PMC;NLO}=\mu_r^{\rm PMC;LO}=\mu_r^{\rm PMC}$, since $\mu_r^{\rm PMC;LO}$ is last known PMC scale. This treatment will lead to some residual scale dependence, which, however, shall be highly exponentially suppressed.

\subsection{The running coupling in the low-energy region}

Eq.(\ref{pmcscale}) indicates that, at the maximum recoil region $q^2=0$, the optimal PMC scale for the $B_c\to \eta_c (J/\psi)$ TFFs is $\mu_r^{\rm PMC}=e^{-\frac{5}{6}}\sqrt{m_b m_c /2}$. If setting $m_b=4.9{\rm GeV}$ and $m_c=1.4{\rm GeV}$, we obtain $\mu_r^{\rm PMC} \approx 0.8{\rm GeV}$, which is close to the low energy (LE) region. In the LE region, the conventional running behavior of the coupling constant may overestimate the pQCD predictions. Several LE effective models have been suggested in the literature~\cite{sjb,Webber,MPT1,MPT2,BPT,CON,GI,AnaMod}. A comparison of six typical LE coupling constant models can be found in Ref.\cite{HQdecay}. The MPT model is phenomenologically successful, i.e. the moment of the spin-dependent structure function calculated within the MPT model is consistent with the experimental data down to a few hundreds of MeV~\cite{MPT2}. For clarity, we shall adopt the MPT model to do our discussion.

The MPT model~\cite{AnaMod,MPT1,MPT2}, based on the massive analytic pQCD theory, provides a convenient way for analyzing the data below $1{\rm GeV}$. It is designed to ensure the nonsingular behavior in the infra-red (IR) region and to eliminate the Landau pole. On the basis of the mass-dependent (massive) Bogoliubov RGE~\cite{mRG}, by introducing the effective gluonic mass $m_{\rm gl} = \sqrt{\xi} \Lambda_{\rm QCD}$ as the IR regulator, $\ln \frac{\mu_r^2}{\Lambda^2_{\rm QCD}} \to \ln \frac{\mu_r^2 + m_{\rm gl}^2}{\Lambda^2_{\rm QCD}}=\ln \left( \xi+\frac{\mu_r^2}{\Lambda^2_{\rm QCD}}\right)$, one can disentangle the unwanted singularity in the IR limit from the usual ultra-violet logs. More explicitly, up to two-loop level, the MPT model suggests
\begin{widetext}
\begin{eqnarray}\label{alphasMPT}
\alpha_{s;\rm MPT}(\mu_r)=\alpha_{\rm crit}\left\{1+\alpha_{\rm crit}\frac{\beta_0}{4\pi} \ln \left(1+\frac{\mu_r^2}{\xi \Lambda^2_{\rm QCD}} \right) + {\alpha_{\rm crit}}\frac{\beta_1}{4\pi \beta_0} \ln \left[1 + \alpha_{\rm crit} \frac{\beta_0}{4\pi} \ln \left(1+\frac{\mu_r^2}
{\xi \Lambda^2_{\rm QCD}} \right) \right] \right\}^{-1},
\end{eqnarray}
\end{widetext}
where $\beta_0=11-\frac{2}{3}{n_f}$ and $\beta_1 = 102 - \frac{38}{3}{n_f}$. The critical running coupling, $\alpha_{\rm crit}=\alpha_{s;\rm MPT}(0)$, is determined via the relation, $\alpha_{\rm crit} ={4\pi}/{({\beta_0}\ln \xi)}$. If setting $\xi=10\pm2$~\cite{MPT2}, we obtain $\alpha_{\rm crit} = 0.606_{-0.044}^{+0.065}$.

\begin{table}[bt]
\centering
\begin{tabular}{c |c|c|c}
 \hline
 & $n_f=3$ & $n_f=4$ & $n_f=5$ \\
 \hline
    ~Conv.~ & $0.388\pm0.007$ & $0.338\pm0.007$ & $0.233\pm0.005$ \\
 \hline
    MPT & $0.260\pm0.005$ & $0.235\pm0.005$ & $0.186\pm0.004$ \\
 \hline
\end{tabular}
\caption{The weighted averages of $\Lambda_{\rm QCD}$ (in unit: GeV) based on the conventional and the MPT $\alpha_s$-running together with the measurements $\alpha_s(M_Z)=0.1185\pm0.0006$ and $\alpha_s(m_\tau)=0.330\pm0.014$~\cite{RPP}. `Conv.' stands for the conventional two-loop $\alpha_s$ running. }  \label{Lambda}
\end{table}

Using the two-loop $\alpha_s$-running together with the measurements $\alpha_s(M_Z)=0.1185\pm0.0006$ and $\alpha_s(m_\tau)=0.330\pm0.014$~\cite{RPP}, the resultant weighted averages for the $\Lambda_{\rm QCD}$ under the conventional and the MPT two-loop $\alpha_s$-running are shown in Table~\ref{Lambda}.

\begin{figure}[hbt]
\includegraphics[width=0.45\textwidth]{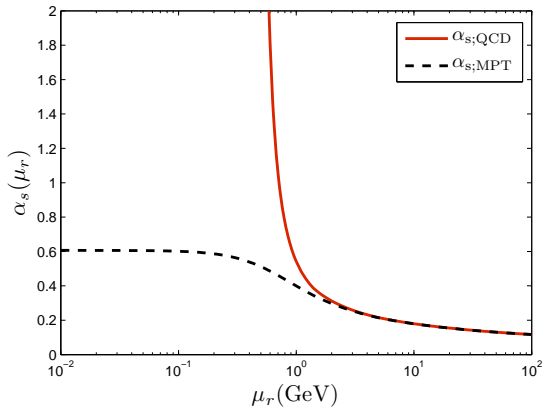}
\caption{A comparison of the strong running coupling $\alpha_s(\mu_r)$ up to two-loop level, where $\alpha_{s;{\rm QCD}}$ stands for the conventional $\alpha_s$ running and $\alpha_{s;{\rm MPT}}$ the MPT-model (\ref{alphasMPT}) with $\xi=10$. }  \label{alphas}
\end{figure}

A comparison of the strong running coupling for the conventional behavior and the MPT-model with $\xi=10$ is put in Fig.(\ref{alphas}). In drawing the curves, the values of $\Lambda_{\rm QCD}$ are taken as their central values shown in Table \ref{Lambda}. It is noted that in the large scale region, both are consistent with each other.

\subsection{Extrapolation of the $B_c \to \eta_c(J/\psi)$ TFFs}

The pQCD predictions for the $B_c \to \eta_c(J/\psi)$ TFFs are reliable in the large recoil region with small $q^2$. In order to explain the semi-leptonic decay, we need the TFFs in all their allowable $q^2$ region. For the purpose, several extrapolations have been suggested, cf.Refs.\cite{Kiselev00,Kiselev02, Huang07,Lu09,BSWFF85,monopole}. In the present paper, we adopt the monopole form, which has been suggested in Ref.\cite{monopole}, to do the extrapolation, i.e.
\begin{eqnarray}
{f_i}(q^2)={f_i}(0)\bigg(1-\frac{q^2}{m_{\rm pole}^2}\bigg)^{-1}, \label{pole}
\end{eqnarray}
where $m_{\rm pole}$ stands for the mass of the lowest-lying resonance. The first derivative of the TFFs over $q^2$ at $q^2=0$, $f'_{i}(0)=f_{i}(0)/m_{\rm pole}^2$, can be calculated within the framework of the QCD sum rules, which inversely can be adopted for determining $m_{\rm pole}$. We take $m_{\rm pole}=4.50$ GeV~\cite{Kiselev00,Kiselev02,qiao13} to do our analysis.

\section{Numerical results and discussions}
\label{sec:3}

We set the $c$- and $b$-quark pole masses as: $m_c = 1.4 \pm 0.1$ GeV and $m_b = 4.9 \pm 0.1$ GeV. And the following PDG values are adopted~\cite{RPP}: $|V_{\rm cb}|=0.0409\pm0.0011$, $m_{\tau}=1.777$ GeV, $m_{Bc}=6.2745$ GeV, $m_{\eta_c}=2.9837$ GeV and $m_{J/\psi}=3.0969$ GeV. We ignore the spin effect for determining the wavefunction at the origin for $\eta_c$ and $J/\psi$, i.e. we adopt $|\psi_{\eta_c}(0)|= |\psi_{J/\psi}(0)|$. The value of $|\psi_{J/\psi}(0)|$ can be determined from the $J/\psi$ leptonic decay width with a relatively high precision. By taking $\Gamma_{J/\psi \to e^+e^-}=5.55\pm0.16$ keV~\cite{RPP} and following the idea of Ref.\cite{wangnpb}, we obtain $|\psi_{J/\psi}(0)|= (0.257_{-0.006}^{+0.010}){\rm GeV}^{3/2}$. As for the wavefunction at the origin for the $B_c$ meson, it can be related with the decay constant via the relation~\cite{psif}, $f_{B_c}^2={12{|\psi_{B_c}(0)|^2}}/{m_{B_c}}$. Using the lattice QCD estimation, $f_{B_c}=(0.489\pm0.005)$ GeV~\cite{fBc}, we obtain $|\psi_{B_c}(0)|=(0.354\pm0.004)\;{\rm GeV}^{3/2}$. As a cross check of our present calculation, when taking the same input parameters as those of Refs.\cite{qiao12,qiao13} , we recover the same numerical results for the $B\to\eta_c(J/\psi)$ TFFs.

\subsection{The TFFs at the maximum recoil region $q^2=0$}

\begin{table}[htb]
\begin{tabular}{c|c|c|c|c}
\hline
 & \multicolumn{3}{c|}{Conventional}  &  PMC \\
\hline
 ~~~$\mu_r^{\rm init}$~~~ & ~~$m_b/2$~~ & ~~$m_b$~~ & ~~$2m_b$~~ & ~~$m_b/2;m_b;2m_b$~~ \\
\hline
$F_1^{B_c\to \eta_c}$ & 1.50 & 1.28 & 1.13 & 1.65 \\
\hline
$F_0^{B_c\to \eta_c}$ & 1.50 & 1.28 & 1.13 & 1.65 \\
\hline
$A_0^{B_c\to J/\psi}$ & 1.08 & 0.97 & 0.89 & 0.87 \\
\hline
$A_1^{B_c\to J/\psi}$ & 1.20 & 1.06 & 0.96 & 1.07 \\
\hline
$A_2^{B_c\to J/\psi}$ & 1.28 & 1.14 & 1.03 & 1.15 \\
\hline
$V^{B_c\to J/\psi}$ & 1.65 & 1.46 & 1.32 & 1.47 \\
\hline
\end{tabular}
\caption{The $B_c \to \eta_c (J/\psi)$ TFFs at $q^2=0$ under the conventional scale setting and the PMC scale setting, in which three typical initial scales are adopted. The PMC scale (\ref{pmcscale}) and hence the PMC predictions are independent of $\mu_r^{\rm init}$. } \label{tab:ffscales}
\end{table}

We put the numerical results for the $B_c \to \eta_c(J/\psi)$ TFFs at the maximum recoil region $q^2=0$ in Table~\ref{tab:ffscales}, where three typical initial scales, $\mu_r^{\rm init}=m_b/2$, $m_b$ and $2m_b$, are adopted. The TFFs decrease with a larger value for $\mu_r^{\rm init}$. Under the conventional scale setting, $\mu_r \equiv \mu_r^{\rm init}$, all the TFFs show a strong scale dependence, i.e. they change by about $[-10\%,+17\%]$ for $\mu_r^{\rm init}\in[m_b/2,2m_b]$. After applying the PMC scale setting, the PMC scale is fixed via Eq.(\ref{pmcscale}), thus, the PMC predictions are independent of the choice of $\mu_r^{\rm init}$.

The usual assumption that the renormalization scale depends on $m_b$ does not have a clear justification. As a byproduct, it is noted that our present scale invariant PMC prediction inversely provides us a chance to set the typical scale for the TFFs. That is, the typical renormalization scale $\mu_r=\mu^{\rm ty}_r$ for the conventional scale setting can be predicted such that it leads to the same TFFs as that of the PMC predictions. Following such argument, we obtain ${\mu_r^{\rm ty}}\simeq 0.3{m_b}$ for $F_{1,0}^{B_c\to \eta_c}$, ${\mu_r^{\rm ty}}\simeq {m_b}$ for $A_{1,2}^{B_c\to J/\psi}$ and $V^{B_c J/\psi}$, and ${\mu_r^{\rm ty}}\simeq 2.3{m_b}$ for $A_{0}^{B_c\to J/\psi}$. This shows that not all of the TFFs have the usual typical scale $m_b$.

\begin{table}[htb]
\begin{tabular}{c|c|c|c|c|c|c}
\hline
 & \multicolumn{3}{c|}{Conventional}  &  \multicolumn{3}{c}{PMC} \\
\hline
 ~~ & ~~LO~~ & ~~NLO~~ & ~~sum~~ & ~~LO~~ & ~~NLO~~ & ~~sum~~ \\
\hline
$F_1^{B_c\to \eta_c}$ & 0.86 & 0.42 & 1.28 & 1.75 & -0.10 & 1.65 \\
\hline
$F_0^{B_c\to \eta_c}$ & 0.86 & 0.42 & 1.28 & 1.75 & -0.10 & 1.65 \\
\hline
$A_0^{B_c\to J/\psi}$ & 0.75 & 0.23 & 0.98 & 1.51 & -0.64 & 0.87 \\
\hline
$A_1^{B_c\to J/\psi}$ & 0.78 & 0.28 & 1.06 & 1.59 & -0.52 & 1.07 \\
\hline
$A_2^{B_c\to J/\psi}$ & 0.84 & 0.30 & 1.14 & 1.70 & -0.55 & 1.15 \\
\hline
$V^{B_c\to J/\psi}$ & 1.08 & 0.38 & 1.46 & 2.19 & -0.72 & 1.47 \\
\hline
\end{tabular}
\caption{The LO and NLO terms for the $B_c\to \eta_c (J/\psi)$ TFFs at $q^2=0$ under the conventional scale setting and the PMC scale setting. $\mu^{\rm init}_r=m_b$. } \label{tab:fflonlo}
\end{table}

After applying the PMC scale setting, due to the elimination of the divergent renormalon terms as $n!\beta^n\alpha_s^n$ with $n$ being the $n$-loop correction, the pQCD convergence can be greatly improved in principle. To show how the pQCD convergence behaves for the $B_c\to \eta_c (J/\psi)$ TFFs, we present the LO and NLO terms for those TFFs at $q^2=0$  before and after the PMC scale setting in Table~\ref{tab:fflonlo}. For clarity, we define a $K$ factor that equals to the magnitude of the ratio between NLO-term and the LO-term, i.e. $K_i=|f_i^{\rm NLO}|/|f_i^{\rm LO}|$. Under the conventional scale setting, for the case of $\mu^{\rm init}_r=m_b$, we obtain
\begin{eqnarray}
&& K_{F_{0}^{B_c\to \eta_c}}=49\%\;,\; K_{F_{1}^{B_c\to \eta_c}}=49\% \;, \nonumber\\
&& K_{A_0^{B_c\to J/\psi}}=31\%\;,\; K_{A_1^{B_c\to J/\psi}}=36\% \;,\nonumber\\
&& K_{A_2^{B_c\to J/\psi}}=36\%\;,\; K_{V^{B_c\to J/\psi}}=35\%\;,
\end{eqnarray}
and after the PMC scale setting, we have
\begin{eqnarray}
&& K_{F_{0}^{B_c\to \eta_c}}=6\%\;,\; K_{F_{1}^{B_c\to \eta_c}}=6\%\;, \nonumber\\
&&K_{A_0^{B_c\to J/\psi}}=42\%\;,\; K_{A_1^{B_c\to J/\psi}}=33\% \;, \nonumber\\
&& K_{A_2^{B_c\to J/\psi}}=32\%\;,\; K_{V^{B_c\to J/\psi}}=33\%\;.
\end{eqnarray}
After the PMC scale setting, the pQCD convergence for the $B_c \to \eta_c$ TFFs $F_{0,1}^{B_c\to \eta_c}$ can be greatly improved, while the $K$ factors for the $B_c \to J/\psi$ TFFs $A_{0,1,2}^{B_c\to J/\psi}$ and $V^{B_c\to J/\psi}$ are still large. The large $K$ factors for the $B_c \to J/\psi$ TFFs indicate that the unknown even higher-order pQCD corrections shall give sizable contributions to the TFFs, which are important either for fixing more precise lower-order PMC scales or for estimating the sizable higher-order conformal contributions. As a minor point, from Table.~\ref{tab:fflonlo}, one may observe that all the NLO corrections to TFFs change from positive values to negative ones after the PMC scale setting.

\begin{table}[htb]
\begin{tabular}{c|c|c}
\hline
 & ~~Conventional~~  &  ~~PMC~~ \\
\hline
$\Delta(F_1^{B_c\to \eta_c})$ & $\pm0.42$ & $\pm0.10$  \\
\hline
$\Delta(F_0^{B_c\to \eta_c})$ & $\pm0.42$ & $\pm0.10$  \\
\hline
$\Delta(A_0^{B_c\to J/\psi})$ & $\pm0.26$ & $\pm0.64$  \\
\hline
$\Delta(A_1^{B_c\to J/\psi})$ & $\pm0.30$ & $\pm0.52$  \\
\hline
$\Delta(A_2^{B_c\to J/\psi})$ & $\pm0.33$ & $\pm0.55$  \\
\hline
$\Delta(V^{B_c\to J/\psi})$   & $\pm0.42$ & $\pm0.72$  \\
\hline
\end{tabular}
\caption{An estimation of the unknown even higher-order pQCD corrections, $\Delta=\pm |{\tilde{\cal C}} \alpha^{2}_s|_{\rm MAX}$, for $B_c\to \eta_c (J/\psi)$ TFFs at $q^2=0$ under the conventional and PMC scale settings. } \label{tab:higherOrder}
\end{table}

For a pQCD estimation, it is helpful to predict what's the ``unknown" QCD corrections could be.  The conventional estimation done by varying the scale over a certain range is not proper, since it can only estimate the non-conformal contribution but not the conformal one. To achieve an estimation of how the ``unknown" QCD corrections could be from the ``known" QCD corrections, a more conservative method for the scale error analysis has been suggested in Ref.\cite{PMC4}; i.e. to take the scale uncertainty as the last known perturbative order. More explicitly, for the present NLO estimation, the pQCD scale uncertainty $\Delta=\pm |{\tilde{\cal C}} \alpha^{2}_s|_{\rm MAX}$, where both ${\tilde{\cal C}}$ and $\alpha_s$ are calculated by varying $\mu^{\rm init}_{r}\in[m_b/2,2m_b]$ and the symbol ``MAX'' stands for the maximum value of $|{\tilde{\cal C}} \alpha^{2}_s|$ within this region. The expression of ${\tilde{\cal C}}$ can be read from Eqs.(\ref{starting},\ref{startingPMC}). We put the $\Delta$ uncertainty for various TFFs in TABLE~\ref{tab:higherOrder}. The large $\Delta$ values also confirm the importance of a next-to-next-to-leading order correction for the $J/\psi$ case. As examples, some PMC analysis up to two-loop, three-loop and four-loop QCD corrections have been done in Refs.\cite{PMC1,PMC2,PMC4,pmcapp1,pmcapp2,pmcapp3,pmcapp4}, which show exactly that the pQCD convergence and the pQCD prediction can be greatly improved after the PMC scale setting.

\subsection{The $B_c\to \eta_c (J/\psi)$ semi-leptonic decays}

\begin{table*}
\begin{tabular}{c |c|c|c|c|c|c|c|c|c|c}
\hline
 & PMC & Conventional & \cite{qiao13} & \cite{nrcqm06} & \cite{rcqm05} & \cite{Huang07,Huang08} & \cite{Chang94,Chang02} & \cite{QCDrpm} & \cite{Kiselev00} & \cite{LOpQCD}\\
 \hline
    $B_c \to \eta_c \ell \nu$ & $71.53_{-8.90}^{+11.27}$ & $43.36_{-5.75}^{+7.17}$ & $30.50^{+9.74}_{-4,82}$ & $6.95^{+0.29}$ & 10.7 & 23.98 & 14.2 & 11.1
    & $11\pm1$ & $6.45^{+1.78}_{-1.59}$ \\
 \hline
    $B_c \to \eta_c \tau \nu$ & $27.14_{-4.33}^{+5.93}$ & $16.46_{-2.73}^{+3.69}$ & $9.29^{+2.70}_{-1.62}$ & $2.46^{+0.07}$ & 3.52 & 7.16 & $\sim$ & $\sim$ & $3.3\pm0.9$ & $2.00_{+0.54}^{-0.50}$ \\
 \hline
    $B_c \to J/\psi \ell \nu$ & $106.31_{-14.01}^{+18.59}$ & $104.74_{-15.49}^{+20.08}$ & $97.30^{+36.22}_{-20.33}$ & $21.9^{+1.2}$ & 28.2 & 34.69 & 34.4 & 30.2 & $28\pm5$ & $14.7^{+1.94}_{-1.73}$ \\
 \hline
    $B_c \to J/\psi \tau \nu$ & $28.25_{-4.35}^{+6.02}$ & $28.12_{-4.72}^{+6.38}$ & $7.55^{+2.85}_{-1.56}$ & $5.86^{+0.23}_{-0.03}$ & 7.82 & 9.50 & $\sim$ & $\sim$ & $7\pm2$ & $4.27^{+0.58}_{-0.50}$ \\
\hline
\end{tabular}
\caption{The decay widths (in unit: $10^{-15}{\rm GeV}$) for the $B_c \to \eta_c (J/\psi) \ell \nu (\ell=e,\mu)$ and $B_c \to \eta_c (J/\psi) \tau \nu$ under the conventional and the PMC scale settings. As a comparison, we also present the results derived from the NLO pQCD factorization~\cite{qiao13}, the constituent quark model~\cite{rcqm05,nrcqm06}, the Bethe-Salpeter equation~\cite{Chang94,Chang02}, the QCD sum rules~\cite{Kiselev00,Huang07,Huang08}, the QCD relativistic potential model~\cite{QCDrpm}, and the LO pQCD~\cite{LOpQCD}.} \label{tab:width}
\end{table*}

\begin{figure}[htb]
\includegraphics[width=0.5\textwidth]{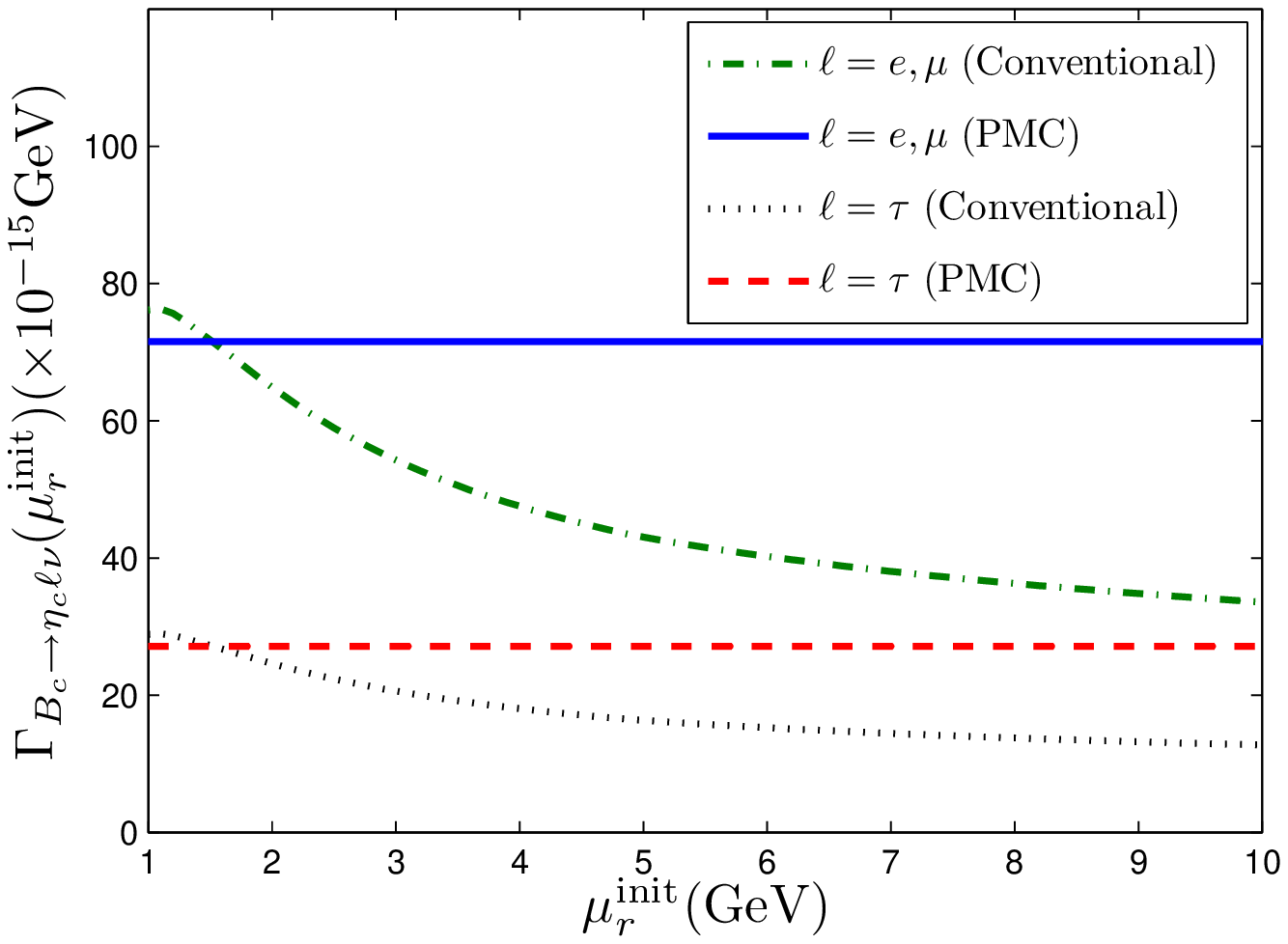}
\includegraphics[width=0.5\textwidth]{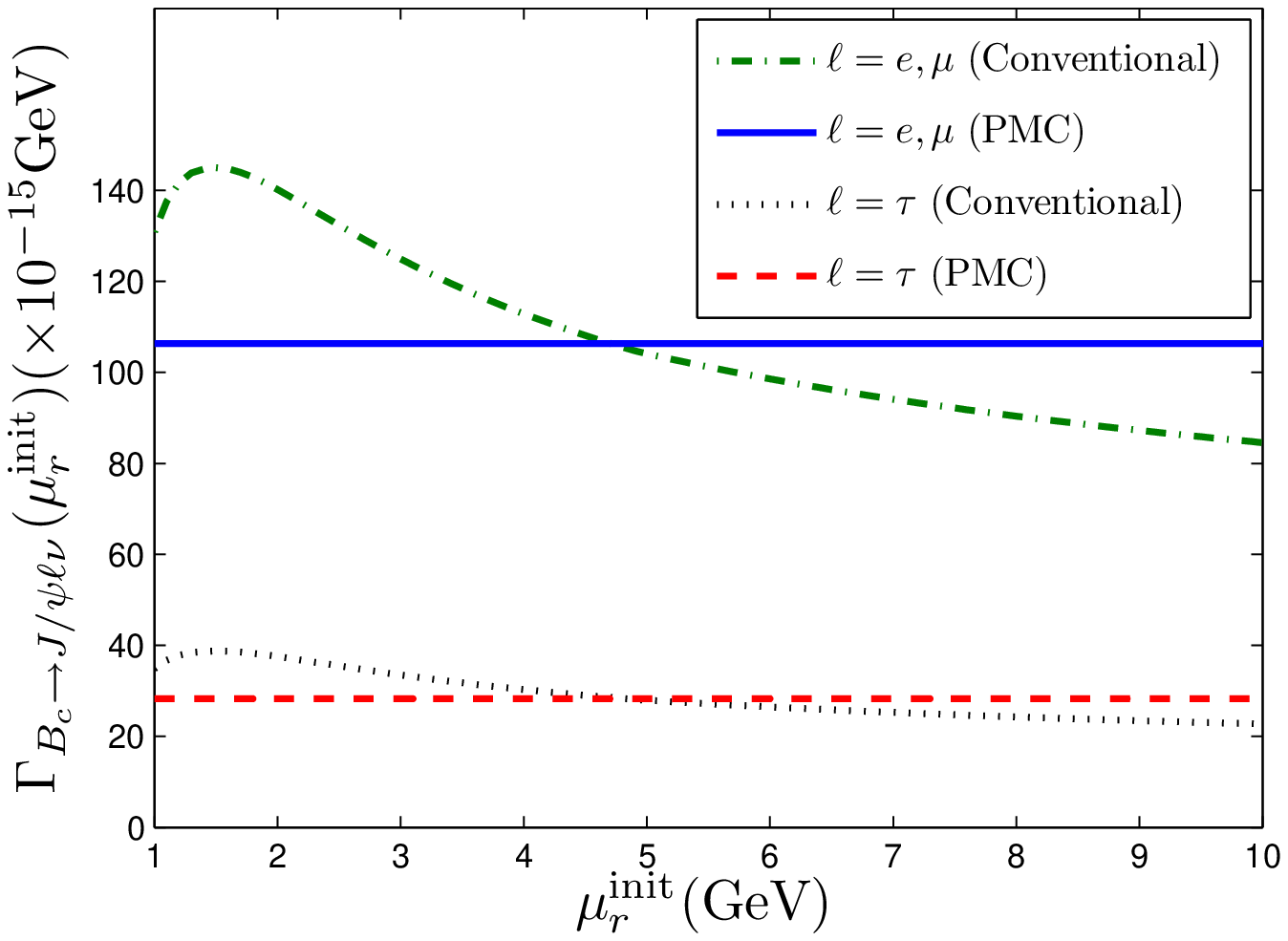}
 \caption{The decay widths for $B_c \to \eta_c(J/\psi) \ell \nu$ up to NLO level versus the initial scale $\mu_r^{\rm init}$ under the conventional and the PMC scale settings, respectively.}
\label{fig:decaywidth}
\end{figure}

By using the monopole extrapolation (\ref{pole}) for the $B_c\to \eta_c (J/\psi)$ TFFs, we are ready to predict the $B_c\to \eta_c (J/\psi)$ semi-leptonic decay widths. The results are presented in Table~\ref{tab:width}, where the errors are the squared average of the mentioned error sources. As a comparison, we present the results before and after the PMC scale setting simultaneously, and we also present the results derived from the NLO pQCD factorization~\cite{qiao13}, the constituent quark model~\cite{rcqm05,nrcqm06}, the Bethe-Salpeter equation~\cite{Chang94,Chang02}, the QCD sum rules~\cite{Kiselev00,Huang07,Huang08}, the QCD relativistic potential model~\cite{QCDrpm}, and the LO pQCD~\cite{LOpQCD}. The renormalization scale dependence of $B_c \to \eta_c(J/\psi) \ell \nu$ decay widths are shown in Fig.(\ref{fig:decaywidth}). After the PMC scale setting, the decay widths are also independent to the choice of $\mu^{\rm init}_r$, which are consistent with the above discussions on the TFFs. Due to the $\tau$ mass suppression, the decay widths for the $\tau$-lepton pair production are smaller than the those of $e$-lepton pair or $\mu$-lepton pair. More specifically, after the PMC scale setting, we have
\begin{eqnarray}
R(\eta_c)&=&\frac{Br(B_c \to \eta_c \ell \nu)}{Br(B_c \to \eta_c \tau \nu)}\simeq2.6, \;{\rm for}\;  \ell=e,\mu,  \\
R(J/\psi)&=&\frac{Br(B_c \to J/\psi \ell \nu)}{Br(B_c \to J/\psi \tau \nu)}\simeq3.8, \;{\rm for}\;  \ell=e,\mu.
\end{eqnarray}

\begin{figure}[htb]
\includegraphics[width=0.45\textwidth]{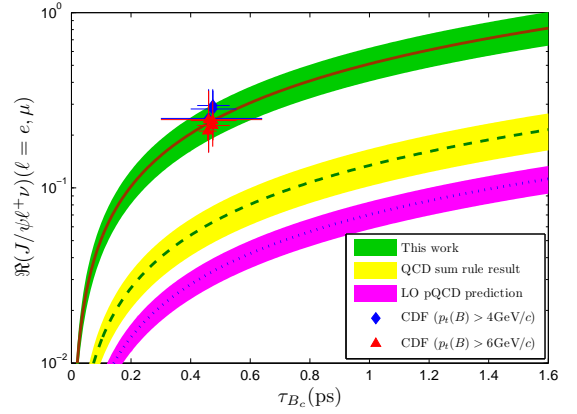}
\caption{The value of $\Re(J/\psi{\ell^+}\nu)$ after the PMC scale setting, which is shown by the upper shaded band. The CDF predictions~\cite{Ru05,Ru09,Ru14,Re05}, the QCD sum rule (SR) prediction~\cite{Kiselev00} and the LO pQCD prediction~\cite{LOpQCD} are presented as a comparison. The middle shaded band represents the QCD sum rule prediction and the lower shaded band is the LO pQCD prediction under the conventional scale setting.}
\label{fig:Rcompare}
\end{figure}

By taking the hadronization fractions $f_{\bar b \to B_c^+}=(1.3\pm0.2)\times10^{-3}$~\cite{Bcprod}, $f_{\bar b \to B^+}=0.401\pm0.008$ and $BR(B^+ \to J/\psi K^+)=(1.028 \pm 0.031)\times 10^{-3}$~\cite{RPP}, we can predict the $\sigma \cdot BR$ ratio $\Re(J/\psi{\ell^+}\nu)$. Our results for $\Re(J/\psi{\ell^+}\nu)$ as a function of the $B_c$ meson lifetime $\tau_{B_c}$ are presented in Fig.(\ref{fig:Rcompare}). As a comparison, the CDF measurements~\cite{Bc98,Ru05,Ru09,Ru14,Re05} as shown in Table~\ref{tab:Rexpe}, where all the errors are added in quadrature. As a comparison, the estimations based on QCD sum rule~\cite{Kiselev00} and LO pQCD prediction~\cite{LOpQCD} are also presented. All those predictions on $\Re(J/\psi{\ell^+}\nu)$ are close in shape, all of which increase with the increment of $\tau_{B_c}$. However our estimation of $\Re(J/\psi{\ell^+}\nu)$ shows a better agreement with the CDF measurements, which indicates the importance of the NLO calculations and also the importance of a correct scale setting.

\subsection{A detailed discussion on the uncertainties of the decay widths}

Table \ref{tab:width} shows the squared average of all errors. In the present subsection, we present a detailed discussion on the dominant error sources. There are many error sources for determining the decay widths, such as the $|V_{cb}|$, $m_{\rm pole}$, $\xi$, $\Lambda_{\rm QCD}$, the bound-state parameters $m_c$, $m_b$, $|\psi_{B_c}(0)|$, $|\psi_{\eta_c}(0)|$ and $|\psi_{J/\psi}(0)|$. Taking the dominant error sources into consideration, we obtain
\begin{eqnarray}
\Gamma_{B_c \to \eta_c \ell \nu} &=&(71.53_{-6.16-3.55-3.80-3.72-0.62}^{+8.19+4.87+3.90+4.55+0.62}) 10^{-15} {\rm GeV} \nonumber\\
&=& (71.53_{-8.90}^{+11.27}) 10^{-15} {\rm GeV},\\
\Gamma_{B_c \to \eta_c \tau \nu} &=&(27.14_{-2.34-3.02-1.44-1.41-0.24}^{+3.11+4.50+1.48+1.73+0.23}) 10^{-15} {\rm GeV} \nonumber\\
&=& (27.14_{-4.33}^{+5.93}) 10^{-15} {\rm GeV},\\
\Gamma_{B_c \to J/\psi \ell \nu} &=&(106.31_{-8.76-8.78-5.64-3.21-0.51}^{+11.74+12.66+5.80+3.68+0.50}) 10^{-15} {\rm GeV} \nonumber\\
&=& (106.31_{-14.01}^{+18.59}) 10^{-15} {\rm GeV},\\
\Gamma_{B_c \to J/\psi \tau \nu} &=&(28.25_{-2.33-3.24-1.50-0.84-0.13}^{+3.12+4.81+1.54+0.96+0.13}) 10^{-15} {\rm GeV} \nonumber\\
&=& (28.25_{-4.35}^{+6.02}) 10^{-15} {\rm GeV},
\end{eqnarray}
where $\ell$ stands for the light leptons $e$ and $\mu$, the uncertainties from the left to right are for a combined effect of the bound state parameters, $m_{\rm pole}$, $|V_{cb}|$, $\xi$, $\Lambda_{\rm QCD}$, respectively. More specifically,
\begin{itemize}
\item The squared average of the uncertainties from the bound state parameters $m_c$, $m_b$, $|\psi_{B_c}(0)|$, $|\psi_{\eta_c}(0)|$ and $|\psi_{J/\psi}(0)|$ are $[-9\%,+11\%]$ for both $B_c \to \eta_c \ell \nu$ and $B_c \to \eta_c \tau \nu$; and $[-8\%,+11\%]$ for both $B_c \to J/\psi \ell \nu$ and $B_c \to J/\psi \tau \nu$.
\item The value of $m_{\rm pole}$ determines the extrapolated shape of the $B_c \to \eta_c(J/\psi)$ TFFs, and we adopt $m_{\rm pole}=(4.50\pm0.25){\rm GeV}$ to do the estimation. The error is $[-5\%,+7\%]$ for $\Gamma_{B_c \to \eta_c \ell \nu}$ and $[-11\%,+17\%]$ for $\Gamma_{B_c \to \eta_c \tau \nu}$; and $[-8\%,+12\%]$ for $\Gamma_{B_c \to J/\psi \ell \nu}$ and $[-11\%,+17\%]$ for $\Gamma_{B_c \to J/\psi \tau \nu}$.
\item The $|V_{cb}|$ being the overall factor for all the $B_c \to \eta_c(J/\psi)$ TFFs, then its error to the decay widths are the same for the channels, which reads $[-5\%,+5\%]$ for $|V_{\rm cb}|=0.0409\pm0.0011$.
\item The errors caused by the MPT parameter $\xi=10\pm2$ are within the region of $[-5\%,+6\%]$ for $\Gamma_{B_c \to \eta_c \ell \nu}$ and $\Gamma_{B_c \to \eta_c \tau \nu}$, and $[-3\%,+3\%]$ for $\Gamma_{B_c \to J/\psi \ell \nu}$ and $\Gamma_{B_c \to J/\psi \tau \nu}$, respectively.
\item By using the values listed in Table \ref{Lambda}, we show that the $\Lambda_{\rm QCD}$ shall cause small errors, i.e. less than $\pm1\%$, for all the decay channels.
\end{itemize}

\begin{figure}[htb]
\includegraphics[width=0.48\textwidth]{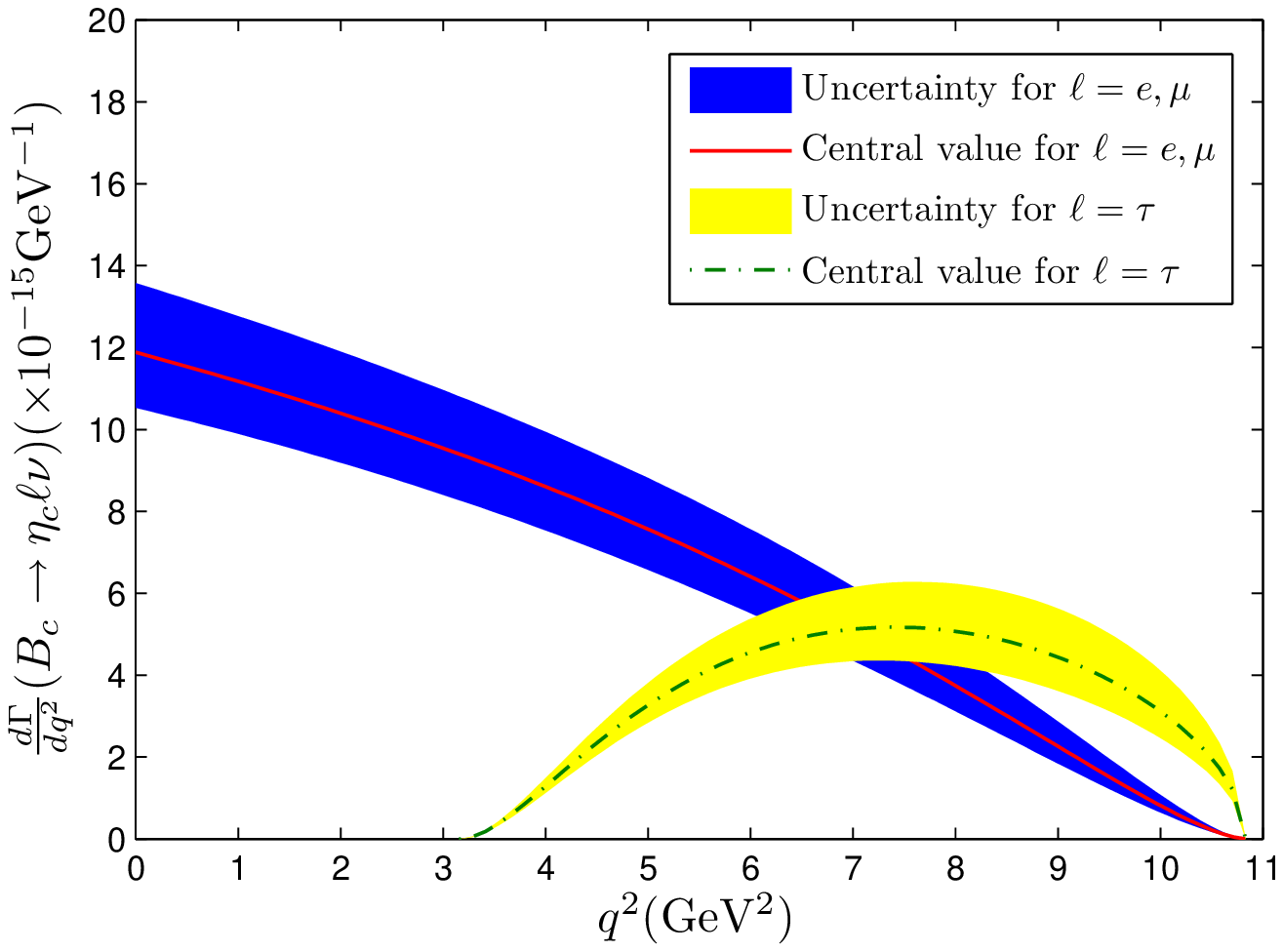}
\includegraphics[width=0.48\textwidth]{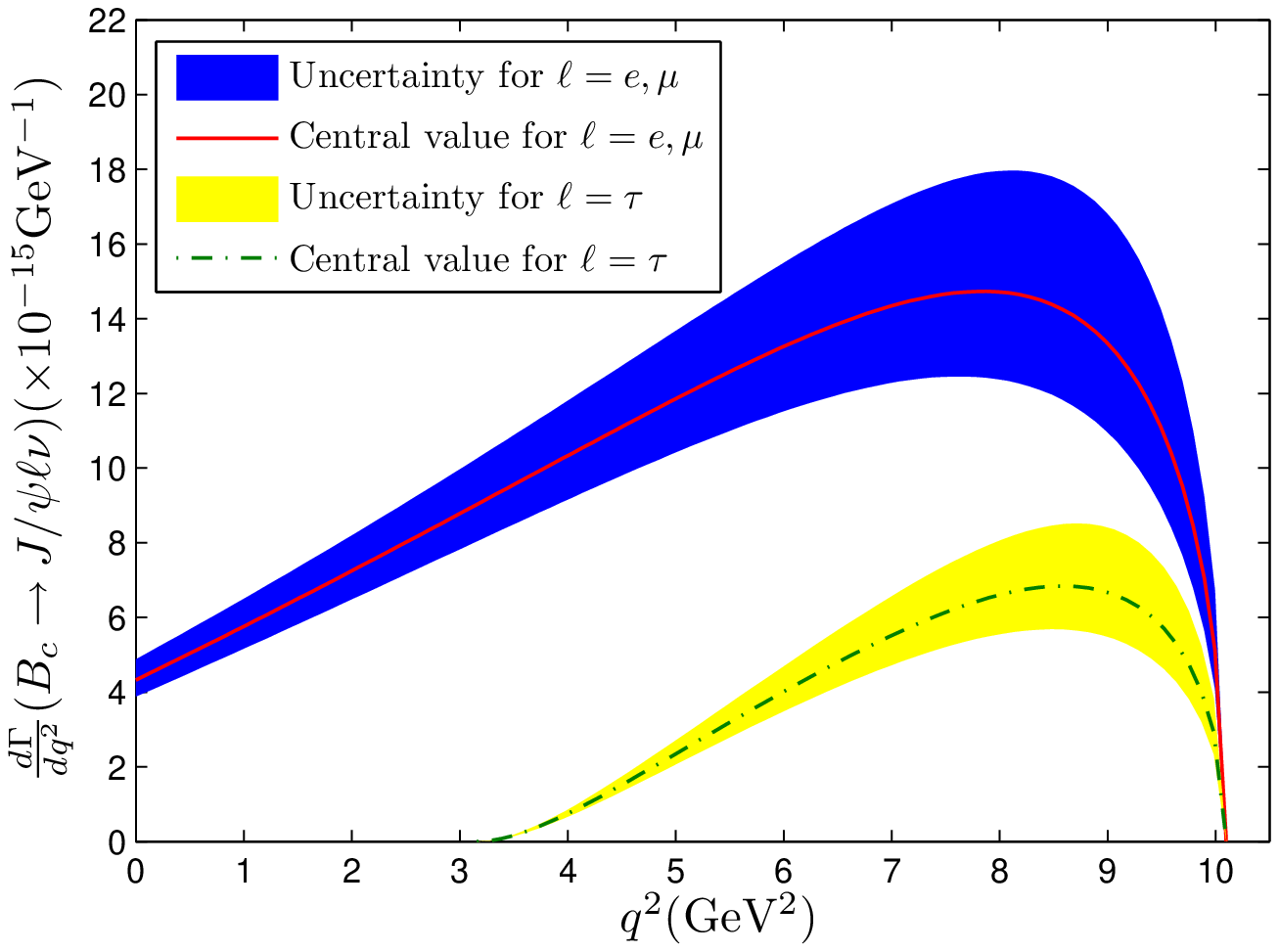} \caption{Differential decay widths for $B_c \to \eta_c(J/\psi) \ell \nu$ $(\ell=e,\mu,\tau)$ versus $q^2$ under the PMC scale setting. }
\label{fig:dgamma}
\end{figure}

It is also helpful to show how those error sources affect the differential decay widths. Also by taking the squared average of those errors, we draw the differential decay widths of $B_c \to \eta_c(J/\psi) \ell \nu(\ell=e,\mu,\tau)$ for $m_{\ell}^2 \leq q^2 \leq (m_{B_c}-m_{\eta_c(J/\psi)})^2$ in Fig.(\ref{fig:dgamma}).

\section{Summary}
\label{sec:4}

The PMC provides a systematic and unambiguous way to set the renormalization scale for any high-energy QCD processes. In the present paper, we have studied the NLO QCD corrections for the $B_c \to \eta_c(J/\psi)$ TFFs by adopting the PMC scale setting. As a further step, we have calculated the $B_c \to \eta_c(J/\psi) \ell \nu$ semi-leptonic decay widths and estimated the measurable parameter $\Re(J/\psi{\ell^+}\nu)$. We have found that
\begin{itemize}
\item After applying the PMC scale setting, all the same type higher-order $\beta_0$-terms have been resummed into the running coupling, which rightly determines the optimal renormalization scale for the $B_c \to \eta_c(J/\psi)$ TFFs. All the $B_{c} \to \eta_c (J/\psi)$ TFFs have the same PMC scale due to the same $\beta_0$-terms at the NLO level. Thus, the large renormalization scale uncertainty for all the TFFs under the conventional scale setting, which is about $[-10\%,+17\%]$ for $\mu_r\in[m_b/2,2m_b]$, are strongly suppressed.

    After applying the PMC scale setting, the pQCD convergence can be improved in principle due to the elimination of the renormalon terms. We have seen an obvious improvement on the pQCD convergence for the $B_c \to \eta_c$ TFFs. However for $B_c \to J/\psi$ TFFs, the $K$ factor is still large, which indicates a NNLO calculation is necessary before an obvious pQCD convergence can be achieved.

\item After applying the PMC scale setting, we obtain larger decay widths for the decays $B_c \to \eta_c(J/\psi) \ell \nu$ in comparison to values under the conventional scale setting, i.e.
    \begin{eqnarray}
    \Gamma_{B_c \to \eta_c \ell \nu}&=&(71.53^{+11.27}_{-8.90})\times 10^{-15} {\rm GeV},\\
    \Gamma_{B_c \to \eta_c \tau \nu}&=&(27.14^{+5.93}_{-4.33})\times 10^{-15} {\rm GeV},\\
    \Gamma_{B_c \to J/\psi \ell \nu}&=&(106.31^{+18.59}_{-14.01})\times 10^{-15} {\rm GeV},\\
    \Gamma_{B_c \to J/\psi \tau \nu}&=&(28.25^{+6.02}_{-4.35})\times 10^{-15} {\rm GeV},
    \end{eqnarray}
    where $\ell=e,\mu$, and the errors are squared averages of those from the dominant sources as $m_c$, $m_b$, $|\psi_{B_c}(0)|$, $|\psi_{J/\psi}(0)|$, $|\psi_{\eta_c}(0)|$, $m_{\rm pole}$, $\xi$, $|V_{cb}|$ and etc..

\item The PMC scale for the decays $B_c \to \eta_c(J/\psi) \ell \nu$ is $\mu_r^{\rm PMC} \approx 0.8{\rm GeV}$, which is in the low-energy region. To provide a reliable pQCD estimation, we have adopted the MPT running coupling model to do the calculation. By taking its input parameter $\xi=10\pm2$, we obtain $[-5\%,+6\%]$ uncertainty for $\Gamma_{B_c \to \eta_c \ell \nu}$ and $\Gamma_{B_c \to \eta_c \tau \nu}$, and $[-3\%,+3\%]$ uncertainty for $\Gamma_{B_c \to J/\psi \ell \nu}$ and $\Gamma_{B_c \to J/\psi \tau \nu}$, respectively.

\item We have estimated the value of $\Re(J/\psi{\ell^+}\nu)$, the production cross section times branching ratio fraction between $B_c^+ \to J/\psi \ell^+ \nu$ and $B^+\to J/\psi K^+$. Our present estimation, as shown in Fig.(\ref{fig:Rcompare}), shows a good agreement with CDF measurements.
\end{itemize}

\hspace{2cm}

{\bf Acknowledgement:}  This work was supported in part by the Fundamental Research Funds for the Central Universities under Grant No.CQDXWL-2012-Z002, by Natural Science Foundation of China under Grant No.11275280.

\hspace{1cm}

\appendix

\section{The LO coefficients and the NLO conformal terms for the $B_c$ to $S$-wave charmonia TFFs at $q^2=0$}

The LO coefficients for the $B_c$ to $S$-wave charmonia TFFs are
\begin{widetext}
\begin{eqnarray}
C^{F_1}(q^2) &=& \frac{8 \sqrt{2} C_A C_F \pi \sqrt{z+1} \left(-\frac{q^2}{m_b^2}+3z^2+2z+3\right)|\psi_{B_c}(0)||\psi_{\eta_c}(0)|}
{\left((1-z)^2-\frac{q^2}{m_b^2}\right)^2 z^{3/2} m_b^3 N_c},
\end{eqnarray}
\begin{eqnarray}
C^{F_0}(q^2) &=& \frac{8 \sqrt{2} C_A C_F \pi \sqrt{z+1}\left(9z^3+9z^2+11z-\frac{q^2}{m_b^2}(5z+3)+3\right) |\psi_{B_c}(0)||\psi_{\eta_c}(0)|}
{\left((1-z)^2-\frac{q^2}{m_b^2}\right)^2 z^{3/2}(3z+1) m_b^3 N_c},
\end{eqnarray}
\begin{eqnarray}
C^V(q^2) &=& \frac{16 \sqrt{2} C_A C_F \pi (3z+1)|\psi_{B_c}(0)||\psi_{\eta_c}(0)|}
{\left( (1-z)^2-\frac{q^2}{m_b^2}\right)^2 \left(\frac{z}{z+1}\right)^{3/2} m_b^3 N_c},
\end{eqnarray}
\begin{eqnarray}
C^{A_0}(q^2) &=& \frac{16 \sqrt{2} C_A C_F \pi (z+1)^{5/2} |\psi_{B_c}(0)||\psi_{\eta_c}(0)|}
{\left((1-z)^2-\frac{q^2}{m_b^2}\right)^2 z^{3/2} m_b^3 N_c},
\end{eqnarray}
\begin{eqnarray}
C^{A_1}(q^2) &=& \frac{16 \sqrt{2} C_A C_F \pi \sqrt{z+1} \left(4z^3+5z^2+6z-\frac{q^2}{m_b^2}(2 z+1)+1\right)|\psi_{B_c}(0)||\psi_{\eta_c}(0)|}
{\left((1-z)^2-\frac{q^2}{m_b^2}\right)^2 z^{3/2} (3 z+1) m_b^3 N_c},
\end{eqnarray}
\begin{eqnarray}
C^{A_2}(q^2) &=& \frac{16 \sqrt{2} C_A C_F \pi \sqrt{z+1} (3z+1) |\psi_{B_c}(0)||\psi_{\eta_c}(0)|}
{\big((1-z)^2-\frac{q^2}{m_b^2}\big)^2 z^{3/2} m_b^3 N_c},
\end{eqnarray}
\end{widetext}
where $C_A=N_c$ and $C_F=(N_c^2-1)/(2N_c)$ with $N_c=3$.

The NLO corrections for all the TFFs have been done in the literature, and the analytic expressions for the TFFs at the $q^2=0$ can be found in Refs.\cite{Bell07,qiao12,qiao13}. For self-consistence and for easy using of the PMC scale setting, we present the NLO conformal terms of the $B_c$-to-Charmonia TFFs in power of $m_c/m_b$ at the maximum recoil region $q^2=0$ in the following:
\begin{widetext}
\begin{eqnarray}
B^{F_1}_{\rm conf}(0) &=& -\frac{55}{12}+\frac{1}{4}\bigg\{-\frac{1}{3} \ln{z}-\frac{2\ln2}{3} +C_F\bigg(\frac{1}{2}\ln^2{z}+\frac{10}{3}\ln2\ln{z}-\frac{35}{6} \ln{z}+\frac{2\ln^2{2}}{3}+3 \ln2+\frac{7\pi^2}{9}-\frac{103}{6}\bigg) \nonumber \\
&&+C_A\bigg(-\frac{1}{6} \ln^2{z}-\frac{1}{3} \ln2 \ln{z} -\frac{1}{3} \ln{z} +\frac{\ln^2{2}}{3}-\frac{4 \ln2}{3}-\frac{5 \pi ^2}{36}+\frac{73}{9}\bigg)\bigg\},
\end{eqnarray}
\begin{eqnarray}
B^{V}_{\rm conf}(0) &=& -\frac{55}{12}+\frac{1}{4}
\bigg\{C_F\bigg(\ln^2{z}+10\ln2\ln{z}-5\ln{z}+9\ln^2{2}+7\ln2+\frac{\pi^2}{3}-15\bigg) \nonumber \\
&&+C_A\bigg(-\frac{1}{2} \ln^2{z}-2\ln2 \ln{z}-\frac{3}{2} \ln{z}-3\ln^2{2}-\frac{3\ln2}{2}-\frac{\pi^2}{3}+\frac{67}{9}\bigg)\bigg\},
\end{eqnarray}
\begin{eqnarray}
B^{A_0}_{\rm conf}(0) &=& -\frac{55}{12}+\frac{1}{4} \bigg\{C_F\bigg(\frac{1}{2} \ln^2{z}-\frac{119}{8}+7 \ln2 \ln{z}-\frac{21}{4} \ln{z}+7\ln^2{2}+\frac{15\ln2}{4}\bigg)\nonumber\\
&&+C_A\bigg(-\frac{3}{8}\ln^2{z}-\ln{2} \ln{z}-\frac{9}{8}\ln{z}-\frac{7\pi^2}{24} +\frac{67}{9}-\frac{9\ln^2{2}}{4}+\frac{3\ln2}{8}\bigg)\bigg\},
\end{eqnarray}
\end{widetext}
where $z=m_c/m_b$. We also have the following relations among the TFFs:
\begin{displaymath}
B^{F_0}_{\rm conf}(0) = B^{F_1}_{\rm conf}(0),\;\; B^{A_1}_{\rm conf}(0) = B^{A_2}_{\rm conf}(0) = B^{V}_{\rm conf}(0) .
\end{displaymath}

\end{document}